\documentclass[amsmath,twocolumn,amssymb,reprint]{revtex4-1}

\usepackage[]{graphicx}
\usepackage{hyperref}
\hypersetup{
    pdfauthor={C. C. Lo},     % author
    pdfcreator={C. C. Lo},   % creator of the document
    pdfnewwindow=true,      % links in new window
    colorlinks=true,       % false: boxed links; true: colored links
    linkcolor=red,          % color of internal links
    citecolor=green,        % color of links to bibliography
    filecolor=magenta,      % color of file links
    urlcolor=cyan           % color of external links
}

\begin{document}
% ========== TITLE ========== % 
\title{Microwave manipulation of electrically injected spin polarized electrons in silicon}
% ======= AUTHOR LIST ====== %
\author{C. C. Lo}
\email{cheuk.lo@ucl.ac.uk.}
\affiliation{London Centre for Nanotechnology, University College London, London WC1H 0AH, U.K.}
\affiliation{Department of Electronic and Electrical Engineering, University College London, London WC1E 7JE, U.K.}
\author{J. Li}
\affiliation{Department of Physics and the Center for Nanophysics and Advanced Materials, University of Maryland, College Park, Maryland 20742, USA}
\author{I. Appelbaum}
\affiliation{Department of Physics and the Center for Nanophysics and Advanced Materials, University of Maryland, College Park, Maryland 20742, USA}
\author{J. J. L. Morton}
\affiliation{London Centre for Nanotechnology, University College London, London WC1H 0AH, U.K.}
\affiliation{Department of Electronic and Electrical Engineering, University College London, London WC1E 7JE, U.K.}
\date{\today}

% ======== ABSTRACT ======== %
\begin{abstract}
We demonstrate microwave manipulation of the spin states of electrically injected spin-polarized electrons in silicon. Although the silicon channel is bounded by ferromagnetic metal films, we show that moderate microwave power can be applied to the devices without altering the device operation significantly. Resonant microwave irradiation is used to induce spin rotation of spin-polarized electrons as they travel across a silicon channel, and the resultant spin polarization is subsequently detected by a ferromagnetic Schottky barrier spin detector. These results demonstrate the potential for combining advanced electron spin resonance techniques to complement the study of semiconductor spintronic devices beyond standard magnetotransport measurements.
\end{abstract}

% ======== PAC CODES AND KEYWORDS ======== %

\maketitle 

% ======== MAIN TEXT ======== %
The spin degree-of-freedom of charge carriers in bulk semiconductors has been studied by electron spin resonance (ESR) for decades, yielding invaluable spin relaxation information~\cite{pifer75,ochiai76}. These measurements demonstrated that the spin relaxation times in group IV semiconductors, notably in silicon, are quite long. Silicon-based spintronic devices are especially relevant due to the widespread use of the material in conventional microelectronics, and hence using silicon as the basis for next generation spintronic devices is even more attractive~\cite{jansen12}. However, while the spin degree of freedom in semiconductor spintronic devices is easily accessible in many group III-V and II-VI compound materials with optical techniques, these approaches are difficult to implement with group IV devices due to the indirect nature of the band gap~\cite{zutic04}, although optical detection of spin injection in silicon has been demonstrated~\cite{Grenet_APL2009}. It is because of this difficulty that most group IV-based semiconductor spintronic devices are studied by quasistatic electrical magnetotransport techniques.  These approaches include the non-local 4-terminal detection of spin accumulation with open-circuit voltage~\cite{johnson88, Lou_NatPhys2007} or spin current with hot electron techniques~\cite{Appelbaum_Nature2007, Huang_PRL2007, Huang_PRB2010, Li_PRL2013}, both of which reveal spin-valve and spin precession effects, and can be used to determine spin-related properties such as spin diffusion length and relaxation times in prototypical spintronic devices. 

\begin{figure} [b]
	\centering
	\includegraphics[width=7.0cm]{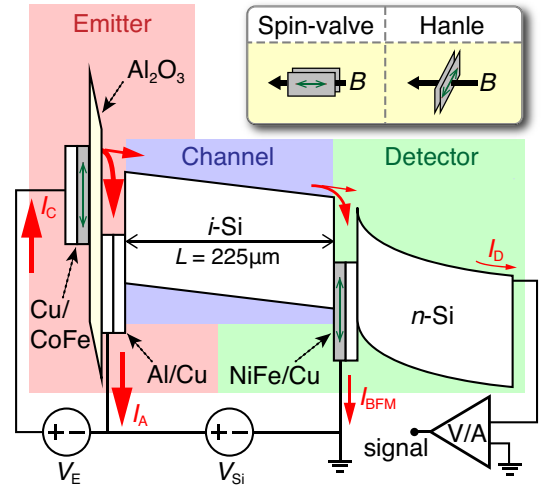}
	\caption{\label{f:1}{Schematics and energy band diagram of the four terminal silicon spintronic device consisting of the emitter, channel and detector. The ferromagnetic layers are shaded in gray. Definitions of bias voltages, voltage polarities and current components (labelled in red) are also indicated. Inset: Orientation of the ferromagnetic layers relative to the applied magnetic field $B$ in the spin-valve and Hanle measurement geometries.}}
\end{figure}

It is thus interesting to consider incorporating electron spin resonance techniques to aid the study of the underlying physics, as advanced pulse sequences will allow the elucidation of the spin dynamics in these devices. We note that ferromagnetic resonance, a closely related technique, has been recently utilized to induce dynamical spin injection (spin pumping) in degenerately doped silicon~\cite{shikoh13}. In the present work, we operate at conditions resonant to the electrons in nearly intrinsic silicon, far away from the ferromagnetic resonances of the metal layers in the device. We show that resonant microwave radiation can be used to induce spin rotation in the silicon channel without severely affecting the device operation, thus paving the way toward pulsed ESR spin manipulation.

% ======== DEVICE DETAILS ======== %
The four-terminal devices studied here consist of two ferromagnetic (FM) thin films sandwiching a silicon (Si) layer in between, forming a spin-valve \cite{Appelbaum_Nature2007}. A schematic of the device and definitions of the voltages and currents are shown in Fig.\:\ref{f:1}. The three components of the device include the emitter, the Si channel and the detector. The emitter consists of a CoFe FM cathode, Al$_2$O$_3$ tunnel barrier and Al/Cu anode for injecting spin-polarized electrons into the channel. The emitter covers an area of approximately $0.02\:$mm$^2$, and is activated by a bias voltage of $V_{\rm E}\lesssim-0.7$\:V reflecting the Schottky barrier height, with the tunnel current ($I_{\rm C}$) increasing quasi-exponentially for more negative voltages. At $V_{\rm E}$ several hundred mV above this threshold, only $\sim$ 0.1\% of the emitter current is ballistically injected into the silicon channel layer, which is nominally undoped Si and has a thickness of $L=225\:\mu$m. By applying different bias voltages across the silicon channel ($V_{\rm Si}$), the average electron transit time to the detector can be varied. The spin detector consists of a buried FM (BFM) layer of NiFe, and a Schottky junction with an $n$-doped (phosphorus) Si substrate. The current measured from the detector ($I_{\rm D}$, ``signal'' in Fig.\:\ref{f:1}) has a magnitude which is $\sim$ 0.1\% of the total injected current. $I_{\rm D}$ depends on the polarization of the electrons reaching the detector due to spin-dependent inelastic scattering in the BFM, and hence it is a measure of spin current. More details of the device fabrication and operation can be found in Ref. \onlinecite{Li_PRL2012}.

All measurements were carried out in an X-band ($\sim\:9.3$\:GHz) ESR spectrometer (Bruker Elexsys 580) with a helium flow cryostat and cylindrical dielectric resonator. The devices were mounted and wire bonded on thin printed circuit boards for establishing electrical connections, and then loaded into the microwave resonator in the cryostat for measurements.  The inset of Fig.\:\ref{f:1} shows the principle measurement configurations used in this study: We will refer to the in-plane magnetic field configuration as the spin-valve geometry, and the out-of-plane configuration as the Hanle geometry.

\begin{figure}
	\centering
	\includegraphics[width=7.5cm, height=19cm]{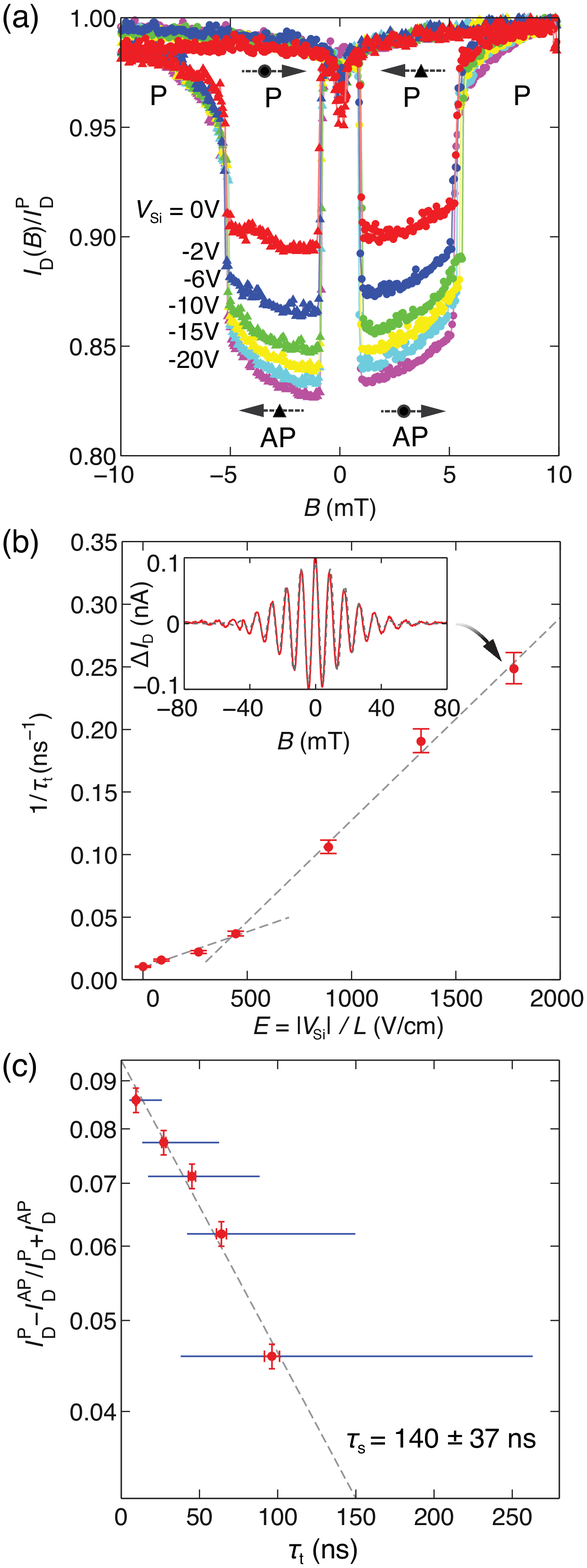}
	\caption{\label{f:2}{(a) Spin-valve measurements carried out at a cryostat temperature of 20\:K. (b, inset) Precession oscillations measured in the Hanle configuration for $V_{\rm Si}=\:-40$\:V (red), with the fitted model shown in dashed lines  (gray), and the extracted peak electron transit times shown in (b). (c) Spin relaxation time $\tau_{\rm s}$ extraction based on the correlation between spin-valve signal amplitude and peak electron transit times. The blue horizontal lines indicate the FWHM of the transit time distribution.}}
\end{figure}

The devices were first characterized by standard magnetotransport measurements to extract the spin flip times ($\tau_{\rm s}$) and carrier transit times ($\tau_{\rm t}$) across the silicon channel. In all the measurements reported we used an emitter bias of $V_{\rm E}=\:-1.5\:$V ($I_{\rm C} \approx -55\:$mA) as it was found to give the best signal-to-noise ratio. Fig.\:\ref{f:2}(a) shows the normalized spin-valve measurements of $I_{\rm D}$ for various $V_{\rm Si}$ at a cryostat temperature of $T=\:20\:$K (the actual device temperature will be discussed shortly). The inset of Fig.\:\ref{f:2}(b) shows precession oscillations observed when monitoring $I_{\rm D}$ in the Hanle configuration. These oscillations unambiguously show that $I_{\rm D}$ does indeed provide a spin-polarization signal~\cite{Huang_PRL2007}, and are well described by a spin drift-diffusion model using current-sensing spin detectors~\cite{appelbaum07b, Li_APL2009}:

\begin{equation}
\label{e:1}
\Delta I_D \propto \int{ \frac{1}{\sqrt{4\pi D t}}\frac{L}{t}e^{-\frac{L^2(1-t/\tau_{\rm t})^2}{4Dt}} e^{-t/\tau_{\rm s}} \cos(\omega t)dt},
\end{equation}

\noindent where $D$ is the electron carrier (and spin) diffusion coefficient and $\omega=g\mu_{\rm B}B/\hbar$ the Larmor precession frequency, with $g$ being the Land$\mathrm{\acute{e}}$ $g$-factor, $\mu_{\rm B}$ the Bohr magneton and $\hbar$ the reduced Planck's constant. The transit time distribution across the channel can be found by examining the Fourier transform of the precession oscillations ~\cite{jang09}. Fig.\:\ref{f:2}(b) shows the peak transit time as the applied electric field, $E=|V_{\rm Si}|/L$, is varied. In a perfectly intrinsic silicon channel where no band bending is expected, $\tau_{\rm t}$ should simply scale inversely with the applied electric field in drift dominated carrier transport. The two distinct regimes observed indicate that residual band bending and diffusive transport plays an increasingly important role in the low electric field regime of $E<500\:$V/cm. We combine the spin-valve and $\tau_{\rm t}$ data to deduce $\tau_{\rm s}$ of electrons in the silicon channel as $ (I_{\rm D}^{\rm P}-I_{\rm D}^{\rm AP})/(I_{\rm D}^{\rm P}+I_{\rm D}^{\rm AP}) \propto e^{-\tau_{\rm t}/\tau_{\rm s}}$, where the superscripts P and AP correspond to the magnetic alignment of the two FM layers in the parallel and anti-parallel configurations, respectively. The results are shown in Fig.\:\ref{f:2}(c), and we obtain $\tau_{\rm s}=\:140\pm37\:$ns. The blue horizontal bars at each data point indicate the FWHM of the transit time distribution, illustrating the significant diffusion-induced broadening of carrier transit times for small drift fields. 

The value of $\tau_{\rm s}$, which is equivalent to the spin-lattice relaxation time $T_1$ in ESR literature, is significantly shorter than would be expected at the cryostat temperature of $T=\:20$\:K~\cite{cheng10}, and indicate heating in the device up to $\approx120\:$K. Sample heating is also confirmed from standard ESR measurements of the background phosphorus signal in the $n$-Si detector, as we found significant reduction in the hyperfine splitting of the phosphorus donor-bound electrons when the emitter bias was increased\cite{digdale71}. In fact, the hyperfine-split donors can no longer be resolved in ESR for $V_{\rm E}\lesssim -1.0$\:V ($I_{\rm C}\approx -16\:$mA), indicating the sample temperature has already risen to above the carrier freeze-out temperature of $\approx 70\:$K. We attribute this large amount of heating to the inefficient thermal anchoring of the sample on the printed circuit board in the flow cryostat, which is especially important when large currents are passed through the emitter for hot electron injection. 

\begin{figure}
	\centering
	\includegraphics[width=7.5cm]{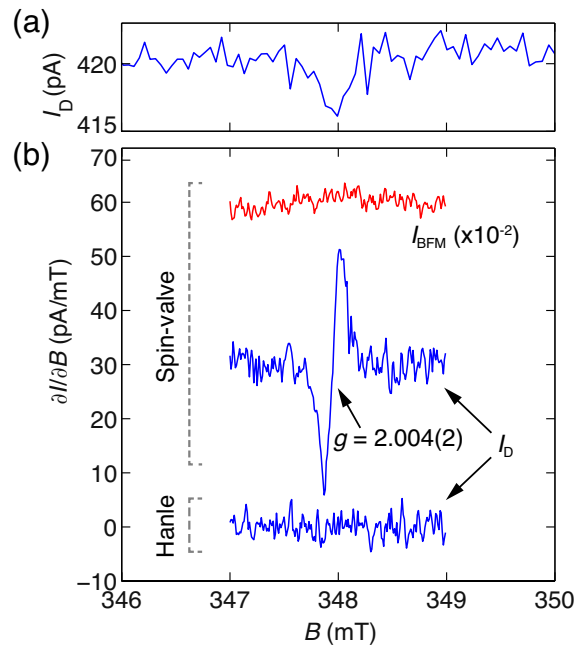}
	\caption{\label{f:3}{Electrically detected spin resonance signal of the silicon spintronic device. (a) Resonant signal detected in the DC current change in detector current $I_{\rm D}$ in the spin valve configuration. (b) Lock-in detection of the buried ferromagnet current $I_{\rm BFM}$ (red) and detector current $I_{\rm D}$ (blue, middle trace) in the spin-valve configuration. When configured in the Hanle geometry, the resonance signal in $I_{\rm D}$ vanishes (blue, bottom trace).}}
\end{figure}

To investigate the effects of resonant microwaves on the device, we turn on and fix the microwave excitation to 9.73\:GHz at 80\:mW, and increase the in-plane magnetic field $B$ to around 350\:mT, which corresponds to the resonance condition for $g\approx 2$ conduction electrons in silicon.  We note that at these magnetic fields the magnetization of the FM layers are always in the parallel configuration, hence $I_{\rm D}$ provides a signal which is proportional to the projection of the spin polarization along the magnetization of the injecting FM. Fig.\:\ref{f:3}(a) shows the variations in $I_{\rm D}$ as $B$ is swept, with $V_{\rm Si}=\:-5\:$V and the sample oriented in the spin-valve configuration. A resonant decrease in $I_{\rm D}$ is observed at approximately 348~mT. 

To confirm that the origin of the resonance signal comes from microwave-induced variations in the electron spin polarization reaching the detector, and not due to other spin-dependent transport mechanisms or bolometric detection in either the Si channel or $n$-Si detector, we examine the orientation and source of spin resonance signal more carefully. We also employ lock-in detection with magnetic field modulation, with $0.2-0.3\:$mT amplitude at 1\:kHz to improve the signal-to-noise ratio. The middle trace in Fig.\:\ref{f:3}(b) shows the lock-in detected signal of $I_{\rm D}$. The current collected by the BFM, shown as the top trace in Fig.\:\ref{f:3}(b), does not reveal any resonance signal at all, implying that the origin of the signal is not due to spin-dependent transport mechanisms in the channel layer, but rather direct spin manipulation by the resonant oscillating magnetic field. 

To further eliminate the possibility that spin-dependent transport processes in the $n$-doped Si detector are responsible for the resonance signal in $I_{\rm D}$, we rotate the sample to the Hanle geometry. In this configuration, the signal disappears as shown in the bottom trace of Fig.\:\ref{f:3}(b), ruling out spin-dependent transport in the $n$-Si or bolometric detection as the source of the resonant signal as no anisotropy is expected from such mechanisms. Indeed, most spin-dependent scattering processes, such as those among donor-conduction electron or conduction electron-conduction electron, are expected to vanish at device temperatures of $T>\:20$\:K ~\cite{lo11, lo11b}. The only strong spin-dependent transport mechanisms in silicon at these temperatures would be spin-dependent hopping or recombination~\cite{brandt04}, neither of which are applicable for the present (undoped) devices. In fact, the vanishing signal in the Hanle geometry is to be expected, since in this case the injected spin polarized electrons are quickly depolarized by the strong out-of-plane magnetic field due to rapid Larmor precession, making the resonant microwave ineffective in inducing any signal variation. These measurements confirm that the resonance signal in $I_{\rm D}$ in the spin-valve configuration is indeed due to microwave-induced rotation of the electron spins in the silicon channel, and its subsequent projection to the BFM magnetization, as detected by $I_{\rm D}$.

We can now examine the effect of the applied electric field on the  resonant change in detector current. The resonant microwave for a given power induces rotation of the electron spins at the Rabi frequency $\omega_1=g\mu_{\rm B}B_1/\hbar$, where $B_1$ is the amplitude of the microwave magnetic field component. As the average $\tau_{\rm t}$ increases (by reducing $E$), so should the electron rotation angle, and hence an increased spin resonance signal. Neglecting spin diffusion effects, we can show that:

\begin{equation}
\frac{\Delta I_D^{\rm P,res}}{I_{\rm D}^{\rm P}-I_{\rm D}^{\rm AP}}\approx \sin^2(\omega_1\tau_{\rm t}/2),
\label{e:1}
\end{equation}

\begin{figure}[t]
	\centering
	\includegraphics[width=7.5cm]{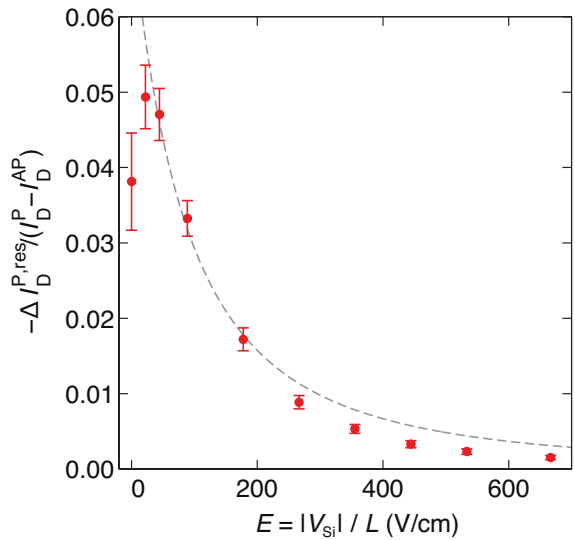}
	\caption{\label{f:4}{Electrically detected spin resonance signal amplitude for various applied electric fields at a cryostat temperature of 20\:K. The dashed line corresponds to fits to the data as described in the main text.}}
\end{figure}

\noindent where $\Delta I_{\rm D}^{\rm P,res}=I_{\rm D}^{\rm P,on-res}-I_{\rm D}^{\rm P,off-res}$ is the integrated electrically detected spin resonance signal amplitude after lock-in detection. Fig.\:\ref{f:4} shows the evolution of $\Delta I_{\rm D}^{\rm P,res}$ normalized in this way for various applied electric fields across the channel. The dashed line is a fit with Eq.\:\ref{e:1} assuming $B_1=\:30\:\mu$T. We note that due to the relatively small $B_1$ field in continuous wave measurements, $\omega_1\approx5\times10^6$\:rad/s, which is slow compared with the transit times achievable in the device, hence no Rabi-like oscillations are expected. In addition, the significant broadening of the transit time distribution (Fig.\:\ref{f:2}(c)) diminishes the achievable oscillation amplitudes. 
We also found increased significance of microwave rectification effects at low bias electric fields; improved sample designs optimized for microwave resonator-based ESR measurements can be used to reduce such detrimental rectification effects~\cite{lo12}, and allow high microwave power pulsed ESR measurements to be performed on the devices.

% ======== Conclusions ======== %
In summary, we have applied standard magnetotransport measurements and combined it with electrical detection of spin resonance to study prototypical silicon spintronic devices. We have shown that the applied spin resonance microwaves can be used to manipulate the injected electron spin states in the silicon transport channel in the spin valve geometry. The vanishing resonance signals in the Hanle geometry clearly rules out other spin-dependent transport mechanisms as the source of the resonance signal. This combination of electron spin resonance and magnetotransport measurements can be an invaluable tool for investigating other spintronic materials and devices, as electron spin resonance techniques can provide precise spectroscopic information and unambiguous interpretation of the temporal spin dynamics in these systems.

% ======== Acknowledgements ======== %
\acknowledgments
Work at UCL is supported by the European Research Council under the European CommunityÕs Seventh Framework Programme (FP7/2007Ð2013)/ERC (grant agreement no. 279781). C.C.L. is supported by the Royal Commission for the Exhibition of 1851. J.L. and I.A. acknowledge the support of the Maryland NanoCenter and its FabLab at U. of Maryland, and funding by the Office of Naval Research under contract N000141110637, the National Science Foundation under contracts ECCS-0901941 and ECCS-1231855, and the Defense Threat Reduction Agency under contract HDTRA1-13-1-0013. J.J.L.M. is supported by the Royal Society.  
% ============================== %


\begin{thebibliography}{10}

\bibitem{pifer75}
J.~H. Pifer.
\newblock Microwave conductivity and conduction-electron spin-resonance
  linewidth of heavily doped si:p and si:as.
\newblock {\em Physical Review B}, 12:4391, 1975.

\bibitem{ochiai76}
Y.~Ochiai and E.~Matsuura.
\newblock Esr in heavily doped n-type silicon near metal-nonmetal transition.
\newblock {\em Phys. Status Solidi A}, 38:243, 1976.

\bibitem{jansen12}
R.~Jansen.
\newblock Silicon spintronics.
\newblock {\em Nature Materials}, 11:400, 2012.

\bibitem{zutic04}
I.~$\mathrm{\check{Z}}$uti$\mathrm{\acute{c}}$, J.~Fabian, and S.~Das Sarma.
\newblock Spintronics: Fundamentals and applications.
\newblock {\em Reviews of Modern Physics}, 76:323, 2004.

\bibitem{Grenet_APL2009}
L.~Grenet, M.~Jamet, P.~Noé, V.~Calvo, J.-M. Hartmann, L.~E. Nistor,
  B.~Rodmacq, S.~Auffret, P.~Warin, and Y.~Samson.
\newblock Spin injection in silicon at zero magnetic field.
\newblock {\em Appl. Phys. Lett.}, 94:032502, 2009.

\bibitem{johnson88}
M.~Johnson and R.~H. Silsbee.
\newblock Spin-injection experiment.
\newblock {\em Physical Review B}, 37(10):5326, 1988.

\bibitem{Lou_NatPhys2007}
Xiaohua Lou, Christoph Adelmann, Scott Crooker, Eric Garlid, Jianjie Zhang,
  K.~Reddy, Soren Flexner, Chris Palmstr{\o}m, and Paul Crowell.
\newblock Electrical detection of spin transport in lateral
  ferromagnet-semiconductor devices.
\newblock {\em Nat. Phys.}, 3:197--202, March 2007.

\bibitem{Appelbaum_Nature2007}
Ian Appelbaum, Biqin Huang, and Douwe Monsma.
\newblock Electronic measurement and control of spin transport in silicon.
\newblock {\em Nature}, 447:295--298, May 2007.

\bibitem{Huang_PRL2007}
B.~Huang, D.~J. Monsma, and I.~Appelbaum.
\newblock Coherent spin transport through a 350 micron thick silicon wafer.
\newblock {\em Physical Review Letters}, 99:177209, 2007.

\bibitem{Huang_PRB2010}
Biqin Huang and Ian Appelbaum.
\newblock Time-of-flight spectroscopy via spin precession: The larmor clock and
  anomalous spin dephasing in silicon.
\newblock {\em Phys. Rev. B}, 82:241202, 2010.

\bibitem{Li_PRL2013}
P.~Li, J.~Li, L.~Qing, H.~Dery, and I.~Appelbaum.
\newblock {\em Phys. Rev. Lett.}, 2013.

\bibitem{shikoh13}
E.~Shikoh, K.~Ando, K.~Kubo, E.~Saitoh, T.~Shinjo, and M.~Shiraishi.
\newblock Spin-pump-induced spin transport in p-type si at room temperature.
\newblock {\em Physical Review Letters}, 110:127201, 2013.

\bibitem{Li_PRL2012}
Jing Li, Lan Qing, Hanan Dery, and Ian Appelbaum.
\newblock Field-induced negative differential spin lifetime in silicon.
\newblock {\em Phys. Rev. Lett.}, 108:157201, 2012.

\bibitem{appelbaum07b}
I.~Appelbaum and D.~J. Monsma.
\newblock Transit-time spin field-effect transistor.
\newblock {\em Applied Physics Letters}, 90:262501, 2007.

\bibitem{Li_APL2009}
Jing Li and Ian Appelbaum.
\newblock Modeling spin transport with current-sensing spin detectors.
\newblock {\em Appl. Phys. Lett.}, 95:152501, 2009.

\bibitem{jang09}
H.~J. Jang and I.~Appelbaum.
\newblock Spin polarized electron transport near the si/sio$_2$ interface.
\newblock {\em Physical Review Letters}, 103:117202, 2009.

\bibitem{cheng10}
J.~L. Cheng, M.~W. Wu, and J.~Fabian.
\newblock Theory of the spin relaxation of conduction electrons in silicon.
\newblock {\em Physical Review Letters}, 104(1):016601, 2010.

\bibitem{digdale71}
D.~E. Dugdale, S.~D. Lacey, and G.~Lancaster.
\newblock A temperature dependent hyperfine interaction in n-type silicon.
\newblock {\em J. Phys. C: Solid St. Phys.}, 4:654, 1971.

\bibitem{lo11}
C.~C. Lo, V.~Lang, R.~E. George, J.~J.~L. Morton, A.~M. Tyryshkin, S.~A. Lyon,
  J.~Bokor, and T.~Schenkel.
\newblock Electrically detected magnetic resonance of neutral donors
  interacting with a two-dimensional electron gas.
\newblock {\em Physical Review Letters}, 106(20):207601, 2011.

\bibitem{lo11b}
C.~C. Lo.
\newblock {\em Electrical detection of spin-dependent transport in silicon}.
\newblock PhD thesis, University of California, Berkeley, 2011.

\bibitem{brandt04}
M.~S. Brandt, S.~T.~B. Goennenwein, T.~Graf, H.~Huebl, S.~Lauterbach, and
  M.~Stutzmann.
\newblock Spin-dependent transport in elemental and compound semiconductors and
  nanostructures.
\newblock {\em Physica Status Solidi (C)}, 1(8):2056, 2004.

\bibitem{lo12}
C.~C. Lo, F.~R. Bradbury, A.~M. Tyryshkin, C.~D. Weis, J.~Bokor, T.~Schenkel,
  and S.~A. Lyon.
\newblock Suppression of microwave rectification effects in electrically
  detected magnetic resonance measurements.
\newblock {\em Applied Physics Letters}, 100(6):063510, 2012.

\end{thebibliography}
\end{document}